\documentclass[%
 aps,
 prb,
 amsmath,
 amssymb,
twocolumn
]{revtex4-1}
\usepackage{amsmath}
\usepackage{bm}

\usepackage{graphicx}
\usepackage{dcolumn}
\usepackage{bm}
\usepackage{color}

\begin{document}

\title{On the interpretation of wave function overlaps in quantum dots}
\author{S. Stobbe}\email[]{ssto@fotonik.dtu.dk}
\author{J. M. Hvam}
\author{P. Lodahl}\homepage[]{www.fotonik.dtu.dk/quantumphotonics}

\affiliation{DTU Fotonik, Department of Photonics Engineering, Technical University of Denmark, {\O}rsteds Plads 343, DK-2800 Kgs.~Lyngby, Denmark}

\date{\today}

\begin{abstract}
The spontaneous emission rate of excitons strongly confined in quantum dots is proportional to the overlap integral of electron and hole envelope wave functions. A common and intuitive interpretation of this result is that the spontaneous emission rate is proportional to the probability that the electron and the hole are located at the same point or region in space, i.e.\ they must coincide spatially to recombine. Here we show that this interpretation is not correct even loosely speaking. By general mathematical considerations we compare the envelope wave function overlap, the exchange overlap integral, and the probability of electrons and holes coinciding and find that the frequency dependence of the envelope wave function overlap integral is very different from that expected from the common interpretation. We show that these theoretical considerations lead to predictions for measurements. We compare our qualitative predictions with recent measurements of the wave function overlap and find good agreement.
\end{abstract}

\maketitle

\section{Introduction}
Self-assembled quantum dots (QDs) are highly interesting for both applications and fundamental studies in many areas of optoelectronics because of their good optical properties and integrability with semiconductor nanotechnology. The latter is a significant advantage over atomic emitters, which implies, e.g., that the optical properties of QDs can be tuned by varying the QD size~\cite{Efros1982,Brus1984,Schmitt-Rink1987,Hanamura1988,Andreani1999}. However, the integration in a solid-state environment can lead to undesired effects such as dephasing~\cite{Gammon1996,Borri1999,Borri2001,Bayer2002} and non-radiative decay processes~\cite{Stobbe2009,Stobbe2010}. Furthermore, the physical understanding of QDs remains much inferior to that of atoms and thus improving the understanding of fundamental concepts of QDs is essential to realize the full potential of QD optoelectronics.

A key optical parameter characterizing an emitter is the oscillator strength (OS), which describes the strength with which the emitter interacts with an electromagnetic field. For an atomic transition the OS has a fixed value, which is in stark contrast to QDs where the OS can be tuned by varying the size of the QD. Coulomb effects are predicted to become dominating for large QDs~\cite{Hanamura1988,Andreani1999}, i.e.\ in the weak-confinement regime, which is relevant also for excitons weakly bound to impurities~\cite{Rashba1962}. For small QDs the Coulomb interaction energy is negligible compared to the energy level spacing originating from the confinement, and in this strong-confinement regime the oscillator strength is proportional to the square of the overlap integral of the envelope wave functions of the electron and the hole~\cite{Efros1982}. The OS of the lowest-energy transition in QDs is typically larger than that of atoms by an order of magnitude and a pronounced frequency dependence of the OS was recently found to be due to the size-dependence of the envelope wave function overlap integral~\cite{Johansen2008}. In this paper we show that very general features of the size dependence of the envelope wave function overlap integral lead to important predictions for the optical properties of QDs in the strong-confinement regime. These predictions are confirmed by both recent experiment and numerical calculations using realistic parameters. Our results show that the common interpretation of the wave function overlap integral being loosely speaking equal to the probability of the electron and hole overlapping spatially is not correct.

\section{Definition of the overlap integrals}
We discuss QDs but note that our results apply for any quantum structure, i.e.\ quantum wires and wells. For QDs the emission energy can be tuned by varying the size and we shall use QD radius and optical angular frequency interchangeably in the understanding that large radii lead to low emission frequencies. The OS $f(\omega)$ of In$_x$Ga$_{1-x}$As QDs in the strong-confinement regime is given by~\cite{Hanamura1988,Stobbe2010}
\begin{equation}
f(\omega) = \frac{E_p(x)}{\hbar\omega} I_\mathrm{WF}(\omega),
\end{equation}
where $E_p(x) = (28.8-7.3x)\:\mathrm{eV}$ is the Kane energy and $\hbar\omega$ is the exciton transition energy. The envelope wave function overlap integral $I_\mathrm{WF}(\omega)$ is defined as
\begin{align}
I_\mathrm{WF}(\omega) &=
\int \mathrm{d}^3\mathbf{r} F_e^\ast(\mathbf{r},\omega) F_h(\mathbf{r},\omega)\nonumber\\
&\times \int \mathrm{d}^3\mathbf{r}' F_h^\ast(\mathbf{r}',\omega) F_e(\mathbf{r}',\omega),\label{Overlap:eq:WFOverlapIntegral}
\end{align}
where $F_e(\mathbf{r},\omega)$ ($F_h(\mathbf{r},\omega)$) is the electron (hole) envelope wave function. We use normalized envelope wave functions, i.e.
\begin{align}
\int \mathrm{d}^3\mathbf{r} \left|F(\mathbf{r},\omega) \right| = 1\label{Overlap:eq:Normalization}
\end{align}
for both electrons and holes, which implies that $F(\mathbf{r},\omega)$ scales as $r^{-3/2}$, where $r$ is the QD radius, which we shall use in the following.

Equation~(\ref{Overlap:eq:WFOverlapIntegral}) is often misunderstood in that it is considered to roughly describe the probability of measuring the electron and the hole in same point or region in space. It is the main purpose of this paper to show that this interpretation is not correct and that it leads to conclusions incompatible with experiments and numerical simulations.

We consider also the exchange integral $I_\mathrm{Ex}(\omega)$, which determines the energy splitting between dark and bright excitons and is of relevance when calculating the spin-flip rate of excitons~\cite{Roszak2007,Roszak2008Erratum,Johansen2008DarkExcitons} and is given by~\cite{Romestain1994}

\begin{align}
I_\mathrm{Ex}(\omega) = \int \mathrm{d}^3\mathbf{r}
\left| F_e(\mathbf{r},\omega) \right|^2 \left| F_h(\mathbf{r},\omega) \right|^2.
\label{Overlap:eq:ExchangeSplittingQD}
\end{align}
It is immediately clear that the exchange integral has the form of an overlap between the probability densities of the electron and the hole. Note, however, that due to Eq.~(\ref{Overlap:eq:Normalization}) $I_\mathrm{Ex}(\omega)$ scales as and has units of inverse QD volume, which means that it cannot be interpreted as a probability and more importantly it has direct consequences for the scaling behavior of $I_\mathrm{Ex}(\omega)$ as discussed below.

Finally, we introduce the quantity describing the probability that both electron and hole are measured in the same volume element $\Omega$, which for instance could be taken to be a unit cell. This quantity is given by the joint probability density, which takes a particularly simple form for independent quantities~\cite{HoelPortStone},
\begin{align}
P_\Omega(\omega) &=
\int_\Omega \mathrm{d}^3\mathbf{r} \left| F_e(\mathbf{r},\omega) \right|^2
\int_\Omega \mathrm{d}^3\mathbf{r}' \left| F_h(\mathbf{r}',\omega) \right|^2.
\end{align}
We can then define the overlap probability as the probability of measuring the electron and hole in the same region of space.
\begin{equation}
P(\omega) = \sum_{i} P_{\Omega_i}(\omega),
\end{equation}
where it is important to realize that $P(\omega)$ depends on $\Omega_i$. Since the squared wave functions are probability distributions the point probabilities vanish and thus $P(\omega)$ vanishes as $\Omega_i$ becomes very small. Thus we must consider finite ${\Omega_i}$.

\section{Size dependence of the overlap integrals}

\begin{figure}
\begin{center}
\includegraphics[width=\columnwidth]{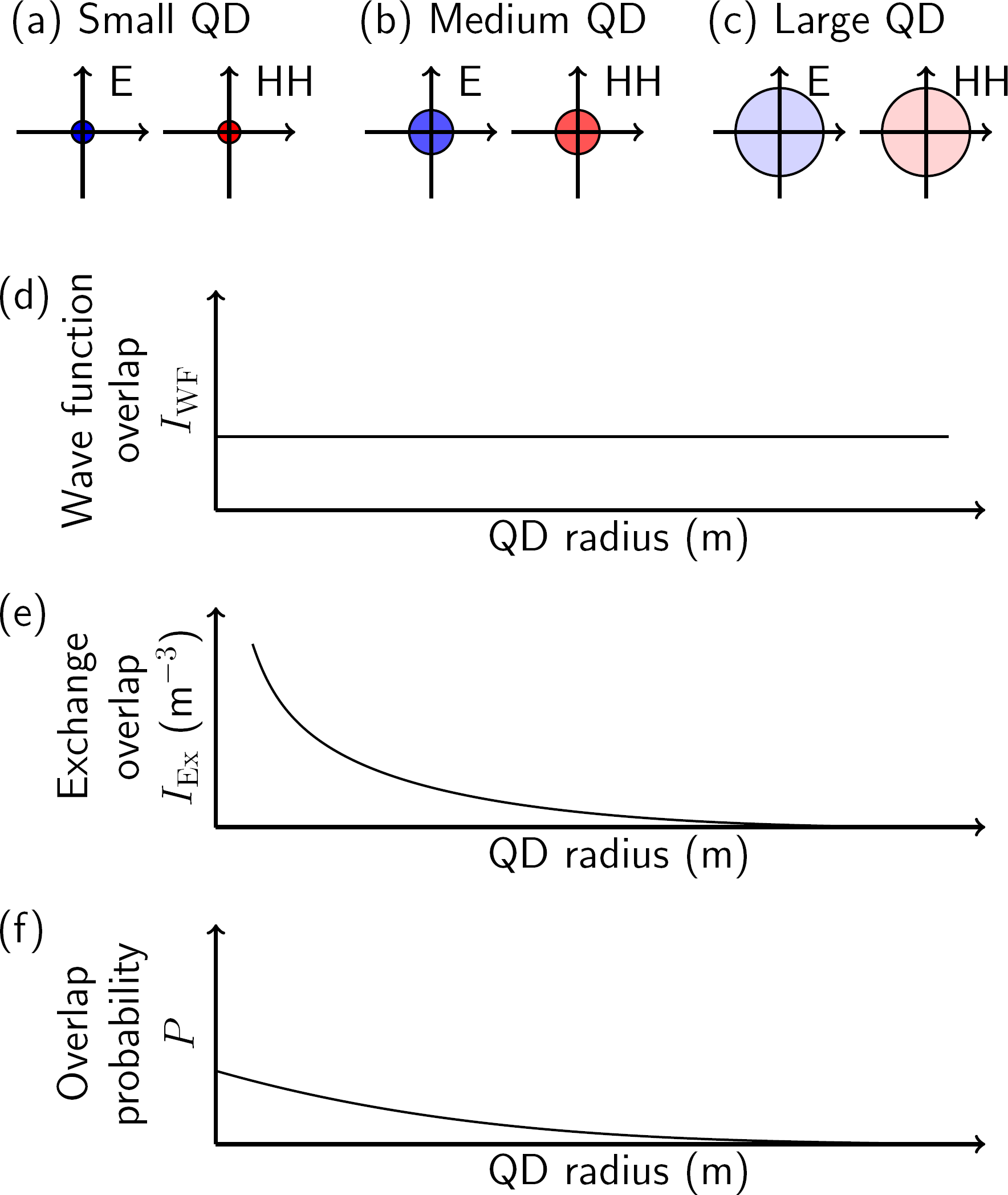}
\caption{Schematic illustration of the size frequency dependence of the overlap integrals for infinite potential barriers in terms of the electron (E) and heavy hole (HH) wave functions. (a)-(c) For infinite barriers the wave functions are independent of the effective mass and hence follow the QD size, which is indicated by the black circles. The wave function amplitude is indicated by the color intensity. This leads to a constant wave function overlap (d) and characteristic size dependencies of the exchange overlap (e) and the overlap probability (f).
\label{Overlap:fig:IntegralComparisonInfinite}}
\end{center}
\end{figure}

\begin{figure}
\begin{center}
\includegraphics[width=\columnwidth]{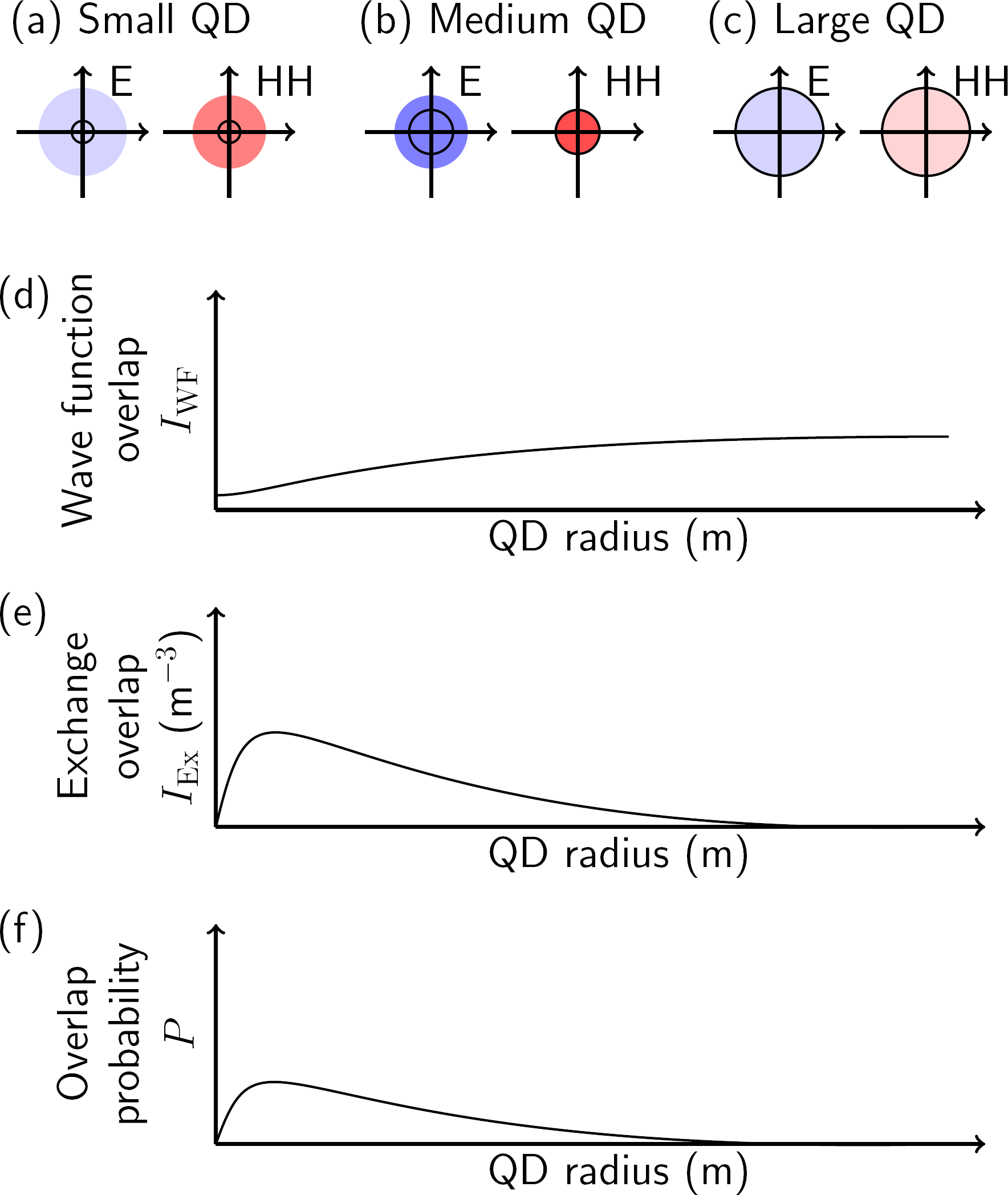}
\caption{The same illustration as Fig.~\ref{Overlap:fig:IntegralComparisonInfinite} but for finite barriers. (a) The wave functions expelled strongly from the QD for very large and very small QDs. (b) In the intermediate regime the lighter electron mass leads to a significant fraction of the electron residing outside the QD while the hole is more confined. (c) For very large QDs the wave functions are approximately identical. This leads to the wave function overlap exhibiting a minimum (d) while the exchange integral (e) and overlap probability (f) exhibit a maximum.
\label{Overlap:fig:IntegralComparisonFinite}}
\end{center}
\end{figure}

The size-dependence of the OS integral is commonly understood as follows: When the size of the QD decreases the electron and hole are gradually expelled from the QD and this is more pronounced for the electrons because of the lighter effective mass as compared to the heavy holes, which are the relevant holes because vertical confinement and strain lift the degeneracy with the light-hole band. The probability of the electron and hole coexisting at the same region in space is reduced and so the oscillator strength must decrease for decreasing QD size, i.e.\ for increasing emission energy. Although this conclusion is correct the above reasoning is in fact incorrect. In the following we discuss this point in further detail.

We consider the limits of very small and very large QD radius as well as the intermediate size regime in order to reconstruct the size dependence and hence frequency dependence of the overlap integrals $I_\mathrm{WF}(\omega)$, $I_\mathrm{P}(\omega)$, and $P(\omega)$. For simplicity, and without loss of generality, we consider a spherical potential. We note that the following considerations are purely mathematical because, e.g., for large radii the strong-confinement model breaks down so that $I_\mathrm{WF}(\omega)$ no longer describes the oscillator strength. Another issue is that for very small radii a spherical three-dimensional potential has no bound states while bound states always exist for one-dimensional potentials. Here we are not concerned with these issues because we are interested in the mathematical limits of the overlap integrals in order to reconstruct their frequency dependences and, as we will show, the mathematical considerations lead directly to physical predictions confirmed by recent experiments.

In Fig.~\ref{Overlap:fig:IntegralComparisonInfinite} we consider infinite potential barriers, which are of little physical relevance but important for the understanding. In this case the electron and hole envelope wave functions are identical and independent of the QD size as indicated in Figs.~\ref{Overlap:fig:IntegralComparisonInfinite}(a)-(c). Therefore the wave function overlap integral is unity and independent of the QD size as shown in Fig.~\ref{Overlap:fig:IntegralComparisonInfinite}(d). The exchange overlap integral scales as inverse volume and therefore it vanishes for very large wave functions and diverges cubically for very small QD radii as shown in Fig.~\ref{Overlap:fig:IntegralComparisonInfinite}(e). For a given $\Omega_i$ the probability of measuring the electron and hole in the same region becomes $P(\omega) = 1$ when the QD radius becomes smaller than $\Omega_i$ while $P(\omega) = 0$ when the QD radius goes to infinity as shown in Fig.~\ref{Overlap:fig:IntegralComparisonInfinite}(f).

Let us now examine the more realistic situation of finite barriers. We can gain physical insight into the problem by considering the length scales entering the system, namely the QD radius and the penetration length $L$ of the wave functions into the surrounding material, which for a one-dimensional square well is given by~\cite{Liboff}
\begin{equation}
L = \frac{\hbar}{\sqrt{2 m^\ast(V_0-E_0)}},
\end{equation}
where $m^\ast$ is the effective mass, $E_0$ is the energy, $V_0$ is the confinement potential, and we have assumed $E_0<V_0$, i.e.\ we are considering a bound state. Here $E_0$ must be found numerically by the wave function continuity criterion, hence $L$ depends on energy and therefore also the QD radius $r$. We can therefore quantify three regimes. In the following $E$ and $HH$ denotes electron and heavy hole respectively. In the small dot regime, i.e.\ $r\ll L_{HH},L_{E}$, the envelope wave functions are strongly expelled from the QD as shown in Fig.~\ref{Overlap:fig:IntegralComparisonFinite}(a). In the intermediate regime, i.e.\ $L_{HH}<r<L_{E}$, the electron is expelled more from the QD than the hole as shown in Fig.~\ref{Overlap:fig:IntegralComparisonFinite}(b). In the large QD regime, i.e.\ $r\gg L_{HH},L_{E}$, the small fractions of the envelope wave functions leaking out of the QD become negligible and the envelope wave functions become effectively identical as shown in Fig.~\ref{Overlap:fig:IntegralComparisonFinite}(c). This size dependence leads directly to the size dependence of the three overlap integrals shown in Figs.~\ref{Overlap:fig:IntegralComparisonFinite}(d)-(f). In the small-dot regime the wave function overlap will have a finite value, which will be very small due to the strong dependence on the difference in effective mass for electrons and holes. The clear qualitative differences between the wave function overlap and the overlap probability in particular for large QD radii directly shows the incorrectness of the common wave function overlap interpretation.

\begin{figure}
\begin{center}
\includegraphics[width=\columnwidth]{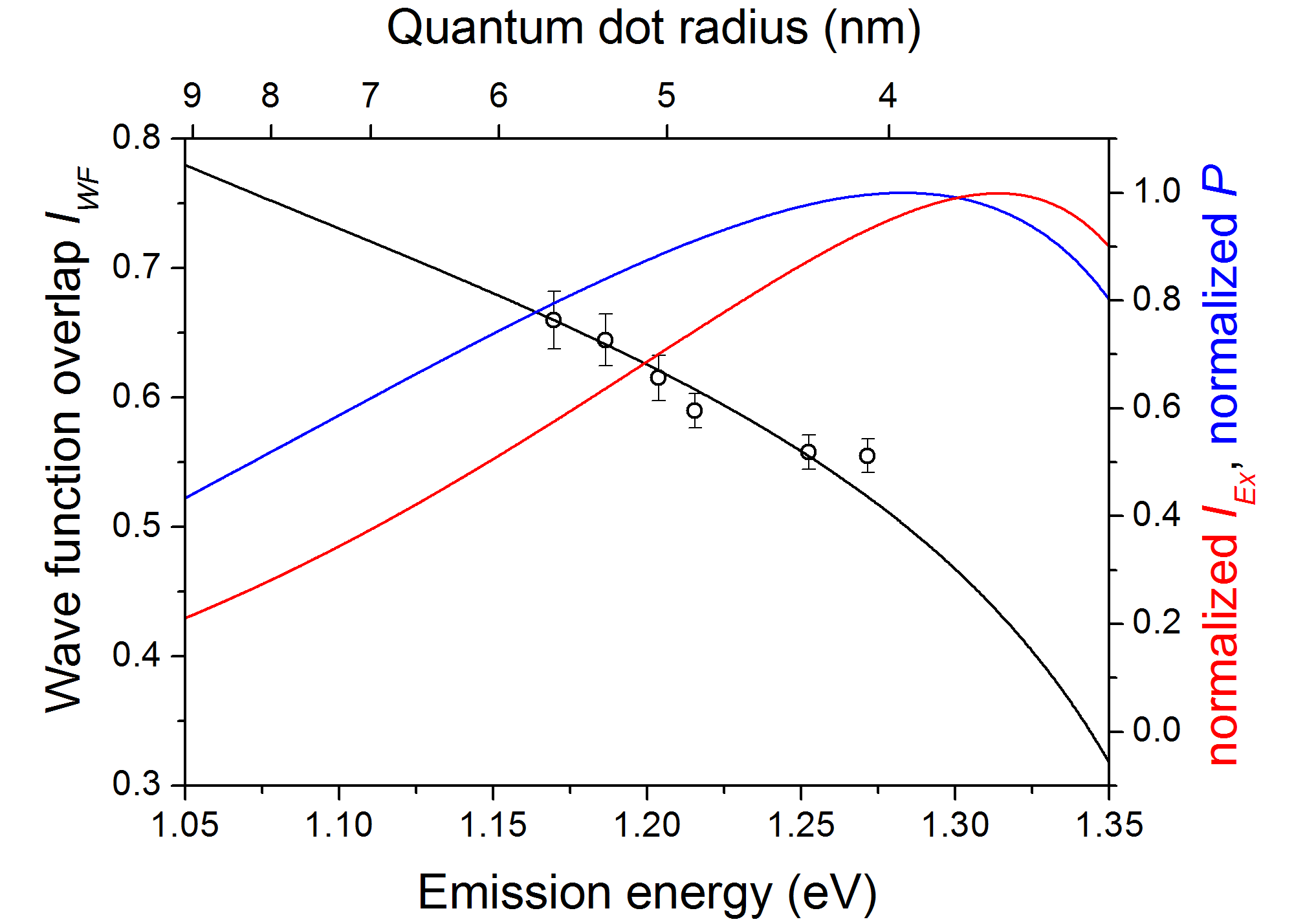}
\caption{Measured wave function overlap (circles) for different emission energies compared to numerically calculated wave function overlap values (black curve, left axis). Using the same numerical parameters we have calculated the exchange overlap (red, right axis) and the overlap probability (blue, right axis), which have been normalized to their maximum values.
\label{Figure3}}
\end{center}
\end{figure}

In order to test the findings above we have performed numerical calculations of the three overlap integrals for a QD comprised of InGaAs with 46\% indium embedded in GaAs using a finite-element model to solve the effective mass equation. The result is shown in Fig.~\ref{Figure3}. The indium mole fraction and size distribution have been optimized to fit recent experimental data on the frequency dependence of $I_{WF}(\omega)$~\cite{Stobbe2009,Johansen2008} as shown in Fig.~\ref{Figure3}. We note, however, that a similar frequency dependence of $I_{WF}(\omega)$ was found for all parameters investigated in Ref.~\cite{Stobbe2009}, i.e.\ also when including various aspect ratios and strain models. Further experimental and numerical details can be found in Ref.~\cite{Stobbe2009}. Notably, as predicted above on general grounds $I_{Ex}(\omega)$ and $P(\omega)$ exhibit very different frequency dependencies from $I_{WF}(\omega)$ in good agreement with the qualitative predictions of the previous sections. In the calculation of $P(\omega)$ we have chosen $\Omega_i \approx 1\:\mathrm{nm}^3$ but the appearance of the maximum remains for all values of $\Omega_i$ that we have investigated.

\section{Conclusion}
In conclusion we have shown that general mathematical aspects of the envelope wave function overlap, the exchange overlap, and the overlap probability lead to distinctly different frequency dependencies. We have confirmed these predictions by numerical calculations and in the case of the envelope wave function overlap also by recent experiments. The essential conclusion is that the envelope wave function overlap must increase with increasing QD size (decreasing emission energy) which is even qualitatively very different from the dependence exhibited by the overlap probability. This shows that the common interpretation of the envelope wave function overlap as an overlap probability is a misconception leading to predictions incompatible with experiments and numerical calculations.

\begin{acknowledgements}
We acknowledge J. Johansen for optical measurements and P. T. Kristensen for valuable discussions. SSt gratefully acknowledges The Danish Council for Independent Research (Project No. FTP 10-080853).
\end{acknowledgements}

\bibliographystyle{pss}

\begin{thebibliography}{[10]}

\bibitem{Efros1982}
 \textsc{A.\,L. \'{E}fros} and  \textsc{A.\,L. \'{E}fros},
 \jr{Soviet physics. Semiconductors} \textbf{16}, 772 (1982).


\bibitem{Brus1984}
 \textsc{L.\,E. Brus},
 \jr{Journal of Chemical Physics} \textbf{80}, 4403 (1984).


\bibitem{Schmitt-Rink1987}
 \textsc{S.~Schmitt-Rink},  \textsc{D.\,A.\,B. Miller},  and  \textsc{D.\,S.
  Chemla},
 \jr{Physical Review B} \textbf{35}, 8113 (1987).


\bibitem{Hanamura1988}
 \textsc{E.~Hanamura},
 \jr{Physical Review B} \textbf{37}, 1273 (1988).


\bibitem{Andreani1999}
 \textsc{L.\,C. Andreani},  \textsc{G.~Panzarini},  and  \textsc{J.\,M.
  G\'{e}rard},
 \jr{Physical Review B} \textbf{60}, 13276 (1999).


\bibitem{Gammon1996}
 \textsc{D.~Gammon},  \textsc{E.\,S. Snow},  \textsc{B.\,V. Shanabrook},
  \textsc{D.\,S. Katzer},  and  \textsc{D.~Park},
 \jr{Science} \textbf{273}, 87 (1996).


\bibitem{Borri1999}
 \textsc{P.~Borri},  \textsc{W.~Langbein},  \textsc{J.~M{\o}rk},
  \textsc{J.\,M. Hvam},  \textsc{F.~Heinrichsdorff},  \textsc{M.\,H. Mao},  and
   \textsc{D.~Bimberg},
 \jr{Physical Review B} \textbf{60}, 7784 (1999).


\bibitem{Borri2001}
 \textsc{P.~Borri},  \textsc{W.~Langbein},  \textsc{S.~Schneider},
  \textsc{U.~Woggon},  \textsc{R.\,L. Sellin},  \textsc{D.~Ouyang},  and
  \textsc{D.~Bimberg},
 \jr{Physical Review Letters} \textbf{87}, 157401 (2001).


\bibitem{Bayer2002}
 \textsc{M.~Bayer} and  \textsc{A.~Forchel},
 \jr{Physical Review B} \textbf{65}, 041308(R) (2002).


\bibitem{Stobbe2009}
 \textsc{S.~Stobbe},  \textsc{J.~Johansen},  \textsc{P.\,T. Kristensen},
  \textsc{J.\,M. Hvam},  and  \textsc{P.~Lodahl},
 \jr{Physical Review B} \textbf{80}, 155307 (2009).


\bibitem{Stobbe2010}
 \textsc{S.~Stobbe},  \textsc{T.~Schlereth},  \textsc{S.~H\"ofling},
  \textsc{A.~Forschel},  \textsc{J.\,M. Hvam},  and  \textsc{P.~Lodahl},
 \jr{Physical Review B} \textbf{82}, 233302 (2010).


\bibitem{Rashba1962}
 \textsc{E.\,I. Rashba} and  \textsc{G.\,E. Gurgenishvili},
 \jr{Soviet Physics. Solid State} \textbf{4}, 759 (1962).


\bibitem{Johansen2008}
 \textsc{J.~Johansen},  \textsc{S.~Stobbe},  \textsc{I.\,S. Nikolaev},
  \textsc{T.~Lund-Hansen},  \textsc{P.\,T. Kristensen},  \textsc{J.\,M. Hvam},
  \textsc{W.\,L. Vos},  and  \textsc{P.~Lodahl},
 \jr{Physical Review B} \textbf{77}, 073303 (2008).


\bibitem{Roszak2007}
 \textsc{K.~Roszak},  \textsc{V.\,M. Axt},  \textsc{T.~Kuhn},  and
  \textsc{P.~Machnikowski},
 \jr{Physical Review B} \textbf{76}, 195324 (2007).


\bibitem{Roszak2008Erratum}
 \textsc{K.~Roszak},  \textsc{V.\,M. Axt},  \textsc{T.~Kuhn},  and
  \textsc{P.~Machnikowski},
 \jr{Physical Review B} \textbf{77}, 249905(E) (2008).


\bibitem{Johansen2008DarkExcitons}
 \textsc{J.~Johansen},  \textsc{B.~Julsgaard},  \textsc{S.~Stobbe},
  \textsc{J.\,M. Hvam},  and  \textsc{P.~Lodahl},
 \jr{Physical Review B} \textbf{81}, 081304(R) (2010).


\bibitem{Romestain1994}
 \textsc{R.~Romestain} and  \textsc{G.~Fishman},
 \jr{Physical Review B} \textbf{49}, 1774 (1994).


\othercit
\bibitem{HoelPortStone}
 \textsc{P.\,G. Hoel},  \textsc{S.\,C. Port},  and  \textsc{C.\,J. Stone},
Introduction to Probability Theory (Houghton Mifflin, 1971).


\othercit
\bibitem{Liboff}
 \textsc{R.\,L. Liboff},
Introductory quantum mechanics (Addison-Wesley, 1998).


\end{thebibliography}
\providecommand{\WileyBibTextsc}{}
\let\textsc\WileyBibTextsc
\providecommand{\othercit}{}
\providecommand{\jr}[1]{#1}
\providecommand{\etal}{~et~al.}

\end{document}